\documentclass{article}
\topskip=\baselineskip
\let\dl=\delta

\let\lm=\lambda

\let\e=\emph

\let\ct=\cite
\let\lf=\left
\let\rt=\right
\let\pa=\partial
\let\la=\langle
\let\ra=\rangle
\let\bv=\mathbf
\let\dt=\cdot
\let\del=\nabla

\let\q=\widehat
\let\Q=\overbrace
\let\h=\hbar

\let\rta=\rightarrow

\let\dy=\displaystyle
\let\ty=\textstyle

\let\hl=\hfill
\let\x=\times
\newcommand{\m}{\mbox}

\newcommand{\ol}[1]{\makebox[\textwidth][s]{#1}}
\newcommand{\id}{\mathrm{I}}
\newcommand{\eqdf}{\stackrel{\mathrm{def}}{=}}
\newcommand{\dlt}{\dl t_{fi}}
\newcommand{\hf}{\ensuremath{{\scriptstyle\frac{1}{2}}}}

\newcommand{\be}{\begin{equation}}
\newcommand{\ee}{\end{equation}}
\newcommand{\dd}[3]{\\ \m{}\\ \ol{\m{#1}\hl\m{${\dy #2}$}\hl\m{#3}}\\ \m{}\\}
\newcommand{\re}[2]{\dd{}{#1}{(#2)}}

\newcommand{\de}[1]{\dd{}{#1}{}}
\newcommand{\ba}{\begin{array}}
\newcommand{\ea}{\end{array}}
\newcommand{\bea}{\begin{eqnarray}}
\newcommand{\eea}{\end{eqnarray}}
\newcommand{\beas}{\begin{eqnarray*}}
\newcommand{\eeas}{\end{eqnarray*}}

\newcommand{\vq}{\bv{q}}
\newcommand{\vv}{\bv{v}}
\newcommand{\va}{\bv{a}}
\newcommand{\vp}{\bv{p}}

\newcommand{\vw}{\bv{w}}

\newcommand{\vR}{\bv{R}}
\newcommand{\vUp}{\bv{\Upsilon}}
\newcommand{\vX}{\bv{X}}
\newcommand{\vk}{\bv{k}}
\newcommand{\vl}{\bv{l}}
\newcommand{\vb}{\bv{b}}
\newcommand{\qvq}{\q{\vq}}
\newcommand{\qvp}{\q{\vp}}
\newcommand{\qq}{\q{q}}
\newcommand{\qp}{\q{p}}
\title{Getting path integrals physically and technically right}
\author{Steven Kenneth Kauffmann \\
        American Physical Society Senior Life Member}
\date{43 Bedok Road \\
      \# 01-11 \\
      Country Park Condominium \\
      Singapore 469564 \\
      Tel \& FAX: +65 6243 6334 \\
      Handphone: +65 9370 6583 \\
      \m{} \\
      and \\
      \m{} \\
      Unit 802, Reflection on the Sea \\
      120 Marine Parade \\
      Coolangatta QLD 4225 \\
      Australia \\
      Mobile:  +61 4 0567 9058 \\
      \m{} \\
      Email: SKKauffmann@gmail.com}
\begin{document}
\maketitle
\begin{abstract}
Feynman's Lagrangian path integral was an outgrowth of Dirac's vague surmise
that Lagrangians have a role in quantum mechanics.  Lagrangians implicitly
incorporate Hamilton's first equation of motion, so their use contravenes the
uncertainty principle, but they are relevant to semiclassical approximations
and relatedly to the ubiquitous case that the Hamiltonian is quadratic in the
canonical momenta, which accounts for the Lagrangian path integral's
``success''.  Feynman also invented the Hamiltonian phase-space path integral,
which is fully compatible with the uncertainty principle.  We recast this as an
\e{ordinary functional integral} by changing direct integration over subpaths
constrained to all have the same two endpoints into an equivalent integration
over those subpaths' unconstrained second derivatives.  Function expansion with
generalized Legendre polynomials of time then enables the functional integral
to be unambiguously evaluated through first order in the elapsed time, yielding
the Schr\"{o}dinger equation with a unique quantization of the classical
Hamiltonian.  Widespread disbelief in that uniqueness stemmed from the mistaken
notion that no subpath can have its two endpoints arbitrarily far separated
when its nonzero elapsed time is made arbitrarily short.  We also obtain the
quantum amplitude for any specified configuration or momentum path, which turns
out to be an ordinary functional integral over, respectively, all momentum or
all configuration paths.  The first of these results is directly compared with
Feynman's mistaken Lagrangian-action hypothesis for such a configuration path
amplitude, with special heed to the case that the Hamiltonian is quadratic in
the canonical momenta.
\end{abstract}

\subsection*{Introduction}

The incorporation of the correspondence principle into quantum
mechanics has proceeded along two profound and elegant parallel
tracks, namely Dirac's canonical commutation rules and Feynman's
path integrals.  It is, however, unfortunately the case that from
their inceptions the \e{prescribed implementations} of both of these
have had some physically unrefined aspects---albeit these conceivable
stumbling blocks turn out to be of little or no \e{practical}
consequence in light of the fact that the Hamiltonians which have been
of interest are almost invariably quadratic forms in the canonical
momenta and as well usually consist of sums of terms which themselves
depend either on \e{only} the canonical coordinates or on \e{only} the
canonical momenta, which makes their unique quantization unmistakably
obvious.  In this paper we nonetheless show that the \e{physically
called-for} refinements of the prescribed implementations of both the
canonical commutation rules and the path integrals result in the
unique quantization of \e{all} classical Hamiltonians rather than
\e{only} those which have heretofore been of practical interest.  This
endows quantum mechanics with a degree of coherence and consistency
which is entirely comparable to that of classical mechanics, and also
renders fully transparent its precise relationship to the latter.

Whereas the called-for refinement of Dirac's canonical commutation rule
prescription is the straightforward strengthening of its classical cor%
respondence to the maximum that is still self-consistent, the physical
issue which besets Feynman's prescribed \e{Lagrangian} path integral
is more drastic.  Because Lagrangians \e{implicitly} incorporate Hamil%
ton's first equation of motion, they likewise \e{implicitly} contravene
the uncertainty principle, which makes their utilization in rigorous
quantum theory impermissible---albeit they \e{do} play a role in semi%
classical approximations and, relatedly, in the practically ubiquitous
special circumstance that the Hamiltonian is a \e{quadratic form in the
canonical momenta}.

In general, however, the \e{Lagrangian} path integral must be regarded
as invalid, and should be replaced by the \e{Hamiltonian phase-space
path integral}, also invented by Feynman, which is fully compatible
with the uncertainty principle.  We recast this as an \e{ordinary
functional integral} by changing direct integration over subpaths con%
strained to all have the same two endpoints into an equivalent inte%
gration over those subpaths' unconstrained second derivatives.  Func%
tion expansion with generalized Legendre polynomials of time then en%
ables the functional integral to be unambiguously evaluated through
first order in the elapsed time, yielding the Schr\"{o}dinger equation
with a unique quantization of the classical Hamiltonian.  Widespread
disbelief in that uniqueness stemmed from \e{misapprehension} of the
fact that \e{arbitrary} endpoint stipulations \e{can always be ful%
filled by an infinite number of subpaths no matter how short the non%
zero time interval allotted for such a subpath may be}.

The unique quantization of the classical Hamiltonian produced by the
Hamiltonian phase-space path integral turns out to be in complete
accord with the unambiguous quantization of that classical Hamiltonian
which emerges from a slightly strengthened, but still self-consistent,
variant of Dirac's canonical commutation rule prescription that is
alluded to above.

This paper also obtains the formal quantum amplitude for a \e{speci%
fied} configuration-space path or a \e{specified} momentum-space path
as an ordinary functional integral over, respectively, \e{all}
momentum-space paths or \e{all} configuration-space paths.  The first
of these two results is then instructively \e{directly} compared and
contrasted with Feynman's mistaken \e{Lagrangian-action hypothesis}
for such a specified configuration-space path amplitude, with special
attention given to the case that the Hamiltonian is a quadratic form
in the canonical momenta.

\subsection*{The Lagrangian path integral}

In the preface to \e{Quantum Mechanics and Path Integrals} by
R. P. Feynman and A. R. Hibbs~\ct{F-H}, which treats \e{only}
the \e{Lagrangian} path integral, the
reader encounters the revelation that, ``Over the succeeding
years, \ldots\ Dr. Feynman's approach to teaching the subject
of quantum mechanics evolved somewhat away from the initial
path integral approach.  At the present time, it appears that
the operator technique is both deeper and more powerful for
the solution of more general quantum-mechanical problems.''
Unfortunately, no recognizable elaboration of this cautionary
note regarding the \e{Lagrangian} path integral is to be found
in the book's main text.  But in what might be construed as a
muffled echo of this theme, we \e{do} learn in the second para%
graph of page 33 of the book that to define the ``normalizing
factor'' $1/A$ which is required to \e{convert} the Dirac-in%
spired very short-time Lagrangian-action phase factor~\ct{Di}%
\ into the \e{actual} very short-time quantum mechanical propa%
gator in configuration representation ``seems to be a very dif%
ficult problem and we do not know how to do it in general
terms''~\ct{F-H}%
.  This makes it clear that the authors, \e{contrary} to a
widely held impression, did \e{not succeed} in making
\e{Lagrangian} path integration into a systematic alternate
approach to quantum mechanics---which one could suppose may
have been reason enough for Feynman to have \e{turned away}
from teaching it.

On page 33 Feynman and Hibbs interpret this ``normalizing factor''
$1/A$ as \e{also} being the ``path measure normalization factor'',
which, when paired with each of multiple integrations over con%
figuration space (at successive, narrowly spaced points in time),
converts the whole lot of those integrations into an actual inte%
gration over \e{all paths} in the limit that the spacing of the
successive time points is taken to zero.  For the \e{particular}
class of one-degree-of-freedom Lagrangians which have the form,
$L(\dot q , q, t) = \hf m{\dot q}^2 - V(q, t)$---to which corres%
ponds the class of quantized Hamiltonians that have the form,
$\q{H}(t) = {\qp}^2/(2m) + V(\qq , t)$---Feynman and Hibbs point
out on page 33 that the factor $1/A$ comes out to equal 
$\sqrt{m/(2\pi i\h\dl t)}$, as that particular quantity properly
converts the $\dl t$-time-interval Lagrangian-action phase factor
into the \e{actual} $\dl t$-time-interval \e{quantum mechanical
propagator in configuration representation}.  Feynman and Hibbs
fail, however, to scrutinize the issue of whether this object can
pass muster as \e{also} being the ``path measure normalization
factor'' which they have, on page 33, \e{explicitly} claimed it
must be.  One notes immediately that this particular $1/A$ depends
on the particle mass $m$, whereas the \e{set} of \e{all} paths could
not possibly depend on anything other than the time interval on
which they are defined and the constraints on their endpoints. The
``measure normalization factor'' for such paths could also feature
constants of mathematics and of nature, but that \e{set} of \e{all}
paths clearly \e{does not change in the slightest} if a
\e{different} value is selected for the particle's \e{mass}!  The
particle mass is a \e{parameter} of the Lagrangian, which is
supposed to be at the heart of the path \e{integrand}---the
\e{measure} aspect of any integral is always supposed to be 
\e{independent} of the choice of \e{integrand}!  Furthermore,
``measure normalization factors'' are, by their nature, supposed
to be \e{positive} numbers, whereas this particular $1/A$ is
complex-valued!  It can only be concluded that the ``Lagrangian
path integral'' simply \e{cannot} make sense as a ``path
integral'' at all!  It is a great pity that Feynman failed to rec%
ognize these \e{surface} anomalies of the ``Lagrangian path integ%
ral'' immediately, as digging deeper only unearths ever worse ones.

Feynman does not seem to have reflected at all on the fact that
mechanical systems that are described by \e{configuration} Lagran%
gians $L(\dot q , q, t)$ can in most instances \e{also} be des%
cribed by \e{momentum} Lagrangians $L(\dot p , p, t)$.  Indeed,
if $L(\dot q , q, t) = \hf m{\dot q}^2 - V(q, t)$, then it turns
out that $L(\dot p , p, t) = -\dot p F^{-1}(\dot p ; t) -
V(F^{-1}(\dot p ; t)) - p^2/(2m)$, where
$F(q; t) \eqdf -\partial V(q,t)/\partial q$.  Unpleasant though
this $L(\dot p , p, t)$ appears for \e{general} $V(q, t)$, it
greatly simplifies when $V(q, t)$ is a \e{quadratic form} in $q$,
e.g., for the harmonic oscillator $V(q, t) = \hf kq^2$,
$L(\dot p , p, t) = {\dot p}^2/(2k) - p^2/(2m)$.  Indeed it will
pretty much be for \e{only} those $V(q, t)$ which \e{are} quadratic
forms in $q$ that the very short-time quantum mechanical propaga%
tor in \e{momentum representation}, which is simply a \e{Fourier
transformation} of the one in configuration representation, will
bear much resemblance to the desired very short-time \e{momentum}
Lagrangian-action phase factor that arises from the quite ugly
$L(\dot p , p, t)$ given above---the good correspondence in the quad%
ratic form cases is an instance of the fact that the Fourier trans%
formation of an exponentiated quadratic form generally comes out
to itself be an exponentiated quadratic form times a simple factor
(albeit that factor is by \e{no means} assured to make sense in
the role of ``path measure normalization factor'', as we have seen
above).  When $V(q, t)$ is \e{not} a quadratic form in q, it will
usually be quite impossible to transparently relate the Fourier
transformation of the very short-time quantum propagator in config%
uration representation to the very short-time Lagrangian-action phase
factor which arises from the fraught $L(\dot p , p, t)$ given above.
The burden of reconciling the two will then have been loaded \e{enti%
rely} onto the shoulders of the $1/A$ factor, whose role as a
\e{fudge factor} will thus have been starkly exposed (its for%
lorn cause as a ``path measure normalization factor'' will certainly
\e{not} have been furthered).

The \e{inability} of the \e{Lagrangian} approach to cope in all but
very fortuitous circumstances with the \e{Fourier transformations}
that take the quantum mechanics configuration representation to
its momentum representation and conversely, suggest a \e{fundamen%
tal incompatability} of Lagrangians with the canonical \e{commuta%
tion rule}, $\qq\qp - \qp\qq = i \h\id$, as that \e{underlies}
the Fourier relation between those representations.  It \e{also},
of course, is the \e{heart} of the \e{uncertainty principle}.
Now Feynman took pains to try to move well away from classical
dynamics by attempting (albeit not so successfully!) to \e{integ%
rate} quantum amplitudes over \e{all} paths, so it does not seem
likely that conflicts with the above quantum canonical commuta%
tion rule could be rooted in that aspect of his approach.  We
have, however, just seen that, aside from Lagrangians of quadra%
tic form, the relationships between $L(\dot q , q, t)$ and
$L(\dot p , p, t)$ exhibit \e{no suggestion of compatibility}
with that commutation rule.  This seems to hint that there may be
something \e{intrinsic to Lagrangians} that is generally incompat%
ible with the quantum momentum-configuration commutation rule.  So
might $L(\dot\vq , \vq , t)$ \e{itself} have a property
that \e{clashes} with the \e{uncertainty principle}?  It turns out
that one need \e{not} look very far to locate \e{that} culprit:
Dirac (and later Feynman) \e{simply failed} to bear in mind
the \e{basic fact} that to \e{any} configuration path
$\vq (t)$, $L(\dot\vq , \vq , t)$ \e{automatically
associates} \e{a uniquely determined momentum path}
$\vp (t) = \del_{\!\dot\vq(t)}L(\dot\vq(t), \vq(t), t)$,
a relation that is \e{patently incompatible} with the
uncertainty principle!

Dirac's vague 1933 surmise about the role of the \e{Lagrangian} in
quantum mechanics~\ct{Di}%
\ has clearly done a long-lived disservice to physics, but Feynman
and also all those who sought to educate themselves in Feynman's
Lagrangian path integral results were as well scientifically
obliged to ponder and pursue any apparently dubious peculiarities
which emanate from them.  H. Bethe blurted out that there are no
paths in quantum mechanics upon hearing Feynman's ideas for the
first time at a Cornell University seminar.  While this initial
visceral reaction cannot be defended as stated, it seems clear
that discomfort concerning the uncertainty principle was
percolating in Bethe's mind.  It is a very great pity that
Bethe did not \e{persist} in pondering that discomfort, seeking
to pin down and confront its source.

\subsection*{The Hamiltonian actions and the phase-space path
integral concept}

Feynman not only originated the Lagrangian path integral idea, he
was also the first to publish the idea of the \e{Hamiltonian phase-%
space path integral}---which he deeply buried in Appendix B of his
major 1951 paper~\ct{Fe}%
.  Apparently he attached little importance to it, and it conceiva%
bly slipped from his mind by 1965, as there is \e{no mention}
of it in the book by Feynman and Hibbs.  Perhaps Feynman had a re%
flexive aversion to all Hamiltonian approaches because of the fact
that the Hamiltonian \e{density} in field theories is \e{not} Lor%
entz-invariant, whereas the Lagrangian density is---that would have
been a pity: the full \e{action density in Hamiltonian form} is
\e{also} a Lorentz invariant; indeed the Lagrangian density is
merely a \e{restricted version} of this.
For quantum theory the Hamiltonian is far superior,
as it does \e{not} harbor the uncertainty principle trap that is
\e{implicit} in the Lagrangian.  To be sure, \e{either one} of the
\e{two} classical Hamiltonian equations of motion \e{does} implici%
tly contradict the uncertainty principle (indeed, the Lagrangian is
a version of the Hamiltonian action integrand that has been \e{re%
stricted} according to \e{one} of the classical Hamiltonian equa%
tions of motion).  But if we firmly drop \e{both} classical Hamil%
tonian equations of motion, $\vq (t)$ and $\vp (t)$ become \e{inde%
pendent} argument functions of the Hamiltonian action functional,
and thus do \e{not} challenge the uncertainty principle.

The path integral concept in this context then becomes one of summing
quantum amplitudes over \e{all phase-space paths}.  This states what
must be done a bit too expansively, however, as we know that in order
to obtain a physically \e{useful} summed amplitude, we must \e{res%
trict} the $\vq(t)$ paths to ones which all have the \e{same} value
$\vq_i$ at the initial time $t_i$ and \e{also} all have the \e{same}
value $\vq_f$ at the final time $t_f$.  An \e{alternate} useful res%
triction is, of course, to require the $\vp(t)$ paths to all have
the \e{same} value $\vp_i$ at the initial time $t_i$ and \e{also} to
all have the \e{same} value $\vp_f$ at the final time $t_f$.  As is
well known, when the configuration paths $\vq(t)$ are endpoint-con%
strained as just described, the two classical Hamiltonian equations
of motion result from setting to zero the first-order variation with
respect to $[\vq(t), \vp(t)]$ of the Hamiltonian action functional,
\re{
S_H([\vq (t), \vp (t)]; t_f, t_i) \eqdf
{\ty\int_{t_i}^{t_f}dt}\,\lf(\dot\vq(t)\dt\vp(t)-H(\vq(t),\vp(t),t)\rt),
}{1a}
whereas when it is the \e{momentum} paths $\vp(t)$ that are endpoint-%
constrained as described above, the \e{same} two classical Hamiltonian
equations of motion result from setting to zero the first-order varia%
tion with respect to $[\vq(t), \vp(t)]$ of the very slightly \e{dif%
ferent} Hamiltonian action functional,
\re{
S_H'([\vq (t), \vp (t)]; t_f, t_i) \eqdf
{\ty\int_{t_i}^{t_f}dt}\,\lf(-\vq (t)\dt\dot\vp(t)-H(\vq(t),\vp(t),t)\rt).
}{1b}
We are, to be sure, interested in \e{summing} the quantum amplitudes
for \e{all} the appropriately endpoint-constrained phase-space paths
rather than in finding \e{which} of those paths is the \e{classical}
one by the variational approach.  Nevertheless, in order to honor
the \e{correspondence principle}, we must make it a path summand
\e{requirement} that the \e{dominant} path as $\h\rta 0$, i.e., the
path of \e{stationary phase}, \e{matches} the classical path. For
that reason, we must be careful to \e{also} match the very slightly
\e{different} actions, $S_H$ or $S_H'$, respectively, to their appro%
priate \e{corresponding} configuration or momentum endpoint con%
straints, respectively, \e{even} in the summands of our \e{path sums}
over \e{quantum amplitudes}---which, in standard fashion, are taken
to be proportional to the exponential of $(i/\h )$ times the \e{ac%
tion} of the path in question.

We also note that that the \e{values} which the two endpoint-%
constraining vectors $\vq_i$ and $\vq_f$ (or, alternately,
$\vp_i$ and $\vp_f$) are \e{permitted} to assume are \e{completely
arbitrary} and \e{mutually independent}.  We shall, in fact,
in quantum mechanical practice frequently be \e{integrating}
over the \e{full range} of either or both of $\vq_i$ and $\vq_f$
(or, alternately, of either or both of $\vp_i$ and $\vp_f$),
so this utter freedom of choice
is, in fact, a \e{necessity}---in the language of quantum
mechanics the \e{range} of both $\vq_i$ and $\vq_f$ (or,
alternately, of both $\vp_i$ and $\vp_f$) must, for \e{each},
describe a \e{complete set} of quantum states.  The
statements just made are neither modified nor qualified
\e{in the slightest} when the \e{positive} quantity
$|t_f - t_i|$ is made \e{increasingly small}.  In other
words, $|\vq_f - \vq_i|$ (or, alternately, $|\vp_f - \vp_i|$)
remains \e{unbounded} no matter \e{how small} the positive
value of $|t_f - t_i|$ may be.  There \e{always exist} an
\e{infinite} number of paths which \e{adhere} to the endpoint
constraints no matter \e{how large} $|\vq_f - \vq_i|$ is or
\e{how small} a positive value $|t_f - t_i|$ assumes.
Indeed, given \e{any velocity} $\vv(t)$ that is
defined for $t\in [t_i, t_f]$ and \e{which satisfies}
$\int_{t_i}^{t_f}dt\,\vv(t) = \vq_f-\vq_i$, the path,
\[\vq(t) = \vq_i+{\ty\int_{t_i}^t}dt'\,\vv(t'),\]
obviously qualifies.  \e{One} such velocity $\vv(t)$ is,
of course, the \e{constant} one, $(\vq_f - \vq_i)/(t_f-t_i)$,
and to it may be added an \e{arbitrary} number of terms of
the form, $\vv^{(n)}(t_i)((t-t_i)^n/n!-(t_f-t_i)^n/(n+1)!)$,
$n=1,2,\ldots\;$.  These \e{utterly elementary} observations
have, in fact, \e{completely eluded the grasp} of an astonishing
number of ``experts'' in the field of path integrals.  Time
and again it is implicitly or explicitly \e{insisted} that,
\[\lim_{|t_f - t_i|\rta 0}|\vq_f - \vq_i|=0,\]
which is then taken to justify the resort to
\e{completely unsound approximations}, in some
instances even a \e{vast class} of these~\ct{Co, Ti}.
This last approach can produce variegated results that
are not merely wrong, but even mutually incompatible!

The endpoint-constraining \e{configuration} vectors $\vq_i$
and $\vq_f$ are, of course, as well part and parcel of the
\e{Lagrangian} path integral, and on their page 38, Feynman
and Hibbs make a variation of the blunder just described.
Their Equation~(2-33) on that page shows a very clear
instance of $\vq_i$ and $\vq_f$ being \e{independently}
integrated, \e{each} over its \e{full range}.  That
\e{notwithstanding}, just below their \e{very next}
Equation~(2-34), they effectively claim that for
sufficiently small $|t_f - t_i|$, the error expression 
$|\vq (t) - \hf (\vq_f + \vq_i)|$ is \e{first-order} in
$|t_f - t_i|$ for all $t$ in the interval $[t_i, t_f]$.
Of course $\vq(t)$ obeys the usual two fundamental
endpoint constraints $\vq (t_i) = \vq_i$ and
$\vq (t_f) = \vq_f$.  These constraints immediately
imply that the above error expression is equal to
$\hf |\vq_f - \vq_i|$ \e{both} at $t = t_i$ \e{and} at
$t = t_f$.  But their \e{independent} integrations over
the \e{full ranges} of $\vq_i$ and $\vq_f$ in their
\e{adjacent} Equation (2-33) make it extremely obvious
that $\hf |\vq_f - \vq_i|$ \e{has no upper bound}!  More%
over, this conclusion is clearly \e{utterly independent}
of \e{how small} a positive value $|t_f - t_i|$ may have!

Having no upper bound is \e{a very long way indeed} from
being first-order in $|t_f - t_i|$ as $|t_f - t_i|\rta 0$!
This massive blunder by the ostensible ultimate
experts in the field drives home the lesson that all
scientists bear the obligation to ponder and pursue
apparently dubious peculiarities \e{irrespective} of their
pedigree.  Science has nothing to gain from the perpetua%
tion of unrecognized mistakes \e{whatever} their source.
The Lagrangian path integral is, of course, deficient
because that approach violates the uncertainty principle,
i.e., it is \e{physically} wrong.  So adding a gross
\e{mathematical} mistake \e{on top} of that doesn't really
much matter.  The critical issue with this particular
category of mathematical blunder is that it has \e{also}
infiltrated the Hamiltonian phase-space path integral,
which has \e{no} known deficiencies of physical principle,
and the \e{manner} of the blunder's intrusion has
\e{completely obfuscated} the unique, straightforward
result which the Hamiltonian path integral in fact yields.

The key consequences of the Hamiltonian phase-space path
integral were first \e{correctly} worked out in a ground%
breaking paper by Kerner and Sutcliffe~\ct{K-S}.  That
paper was quickly taken to task by L. Cohen~\ct{Co}
\e{because it failed to take into account the full
consequences of the ``fact'' that}
$\lim_{|t_f - t_i|\rta 0}|\vq_f - \vq_i|=0$!  Cohen's
``fact'' is, of course, as we have gone to great
pains above to demonstrate, a \e{baneful fiction}!
A consequence of the toxic assumption that
$\lim_{|t_f - t_i|\rta 0}|\vq_f - \vq_i|=0$ is, according
to Cohen and his followers Tirapegui et al.~\ct{Ti},
that for all sufficiently small positive values of
$|t_f - t_i|$, the term $H(\vq (t), \vp (t), t)$ in the
integrand of the Hamiltonian action in Eq. (1a) may, for
\e{all} $t$ in the interval $[t_i, t_f]$, always be
replaced by \e{any constant-in-time} ``discretization''
entity of the form $h(\vq_f, \vq_i, \bar\vp , \bar t)$,
where $\bar\vp$ can be regarded as a type of average
value of $\vp (t)$ for t in the interval $[t_i, t_f]$,
$\bar t$ is some \e{fixed} element of that interval, and
$h$ is \e{any} smooth function that satisfies
$h(\vq ,\vq ,\vp , t) = H(\vq ,\vp , t)$.  Thus,
$H(\hf (\vq_f + \vq_i), \bar\vp , \bar t)$---which is
effectively the same as the \e{bad approximation} to
$\vq (t)$ by $\hf (\vq_f + \vq_i)$ advocated by Feynman
and Hibbs---is one such ``discretization''.  The quasi-%
\e{optimized} ``discretization''
$\hf (H(\vq_f, \bar\vp , \bar t) + H(\vq_i, \bar\vp , \bar t))$
is nonetheless \e{also} a \e{bad approximation}, as
can be verified by examining its differences from
$H(\vq (t), \vp (t), t)$ at the two endpoints $t = t_i$
and $t = t_f$ when $\vq_f$ is assumed to be \e{arbitrar%
ily different} from $\vq_i$.  The remaining members of
this \e{vast} class of ``discretizations'' are \e{bad
approximations} as well, as similar arguments about how
badly they can miss at one or the other or both of those
two time endpoints shows.  One upshot of the
\e{misguided imposition} of this vast ``discretization''
class of \e{unsound approximations} on the Hamiltonian
phase-space path integral is to foster the \e{false im%
pression} that the Hamiltonian path integral does \e{not}
yield a \e{unique} result---indeed that it even paradox%
ically simultaneously yields quite a few \e{mutually
incompatible} results!  The \e{correct} treatment of
this path integral \e{does} in fact yield a unique
result; it is merely the fact that \e{different} members
of this \e{vast class} of \e{unsound} ``discretization''
\e{approximations} can \e{differ substantially from each
other} that lies behind the pedestrian phenomenon that
two \e{different} unsound ``discretization'' approxima%
tions can produce two \e{sufficiently different} wrong
results such that they are in fact \e{mutually incompat%
ible}.  Tirapegui et al.~\ct{Ti} actually set to work
\e{categorizing} this vast class of \e{unsound}
``discretization'' approximations and their frequently
mutually incompatible results---\e{all} of which are,
in fact, nothing more than the \e{counterproductive
fruit} of Cohen's \e{completely erroneous assertion}
that $\lim_{|t_f - t_i|\rta 0}|\vq_f - \vq_i|=0$!

\subsection*{Expressing the Hamiltonian phase-space path integral
             in efficacious form}

With the burden of Cohen's counterproductive mathematical
lapse---which has been permitted to block understanding
for far too many decades---lifted, we turn our attention
to trying to express the Hamiltonian phase-space path
integral in a form that is as understandable and effica%
cious technically as it is physically.  We wish to make
the concept of summing quantum amplitudes over all phase-%
space paths \e{completely central} ab initio, \e{not} to
\e{stumble} on it in conseqence of first having written
down a great many repeated integrations over configuration
(or momentum) space that arise from some \e{other} approach
to quantum mechanics.  This strategy automatically orients
our thinking toward the concept of \e{functional integration}.
However, a stumbling block to the immediate identification
of the phase-space path integral as simply a functional
integral arises from the physically key but technically awk%
ward set-of-measure-zero \e{endpoint constraints} on the
configuration (or momentum) paths.  Because of this obstacle
we \e{begin} by writing the configuration version of the
phase-space path integral in the standard merely
\e{schematic} form, with those problematic endpoint con%
straints \e{only} expressed in words,
\re{K_H(\vq_f, t_f; \vq_i, t_i) =
{\ty\int}\mathcal{D}^{(t\in [t_i, t_f])}_{[\vq (t), \vp (t)]}\,
\exp(iS_H([\vq (t), \vp (t)]; t_f, t_i)/\h ),
}{2}
where it is \e{understood} that all the $\vq (t)$ configura%
tion paths that enter into the ``functional integral'' on the
right-hand side of Eq.~(2) are \e{restricted} by the endpoint
constraints $\vq (t_i) = \vq_i$ and $\vq (t_f) = \vq_f$.  It
now behooves us, of course, to discover mathematical machinery
which gives \e{actual effect} to that understanding!  Before
getting to grips with this issue, however, we note that the
configuration path integral $K_H(\vq_f, t_f; \vq_i, t_i)$ is
to be given its usual quantum-mechanical interpretation as the
time-evolution operator for the wave function in configuration
representation, i.e.,
\re{\psi(\vq_f, t_f) =
{\ty\int}K_H(\vq_f, t_f; \vq_i, t_i)\psi(\vq_i, t_i)d^n\vq_i,
}{3a}
which requires, inter alia, that,
\re{\lim_{t_f\rta t_i}K_H(\vq_f, t_f; \vq_i, t_i) =
\dl^{(n)}\!(\vq_f - \vq_i).
}{3b}
\indent
Eq.~(3a) certainly underlines the critical quantum-mech%
anical role which the two endpoint constraints $\vq (t_i)
= \vq_i$ and $\vq (t_f) = \vq_f$ play in the schematic
Eq.~(2) path integral.  But these endpoint constraints
would have little practical effect if the $\vq(t)$ paths
were not \e{simultaneously} required to be \e{sufficiently
smooth} for all $t\in[t_i, t_f]$.  If the $\vq(t)$ paths were
permitted to have \e{jump discontinuities}, for example, it is
obvious that the two endpoint requirements would insignificantly
constrain them.  Now in \e{classical} mechanics one \e{varies}
rather than \e{sums} over the phase-space $(\vq(t), \vp(t))$
paths, and since the classical path obeys a differential
equation that is second-order in time, it suffices to vary
\e{only} over phase-space paths which are continuously twice
differentiable.  We shall therefore impose exactly this smooth%
ness requirement on the phase-space paths that we \e{sum} over,
i.e., we \e{only} sum over phase-space paths which are \e{smooth
enough} to be classical path \e{candidates}.  In this regard, we
take note of the fact that given \e{any} continuous acceleration
$\va(t)$ defined for $t\in[t_i, t_f]$, there exists a continuously
twice-differentiable $\vq(t)$ defined for $t\in[t_i, t_f]$ which
satisfies the three conditions $\vq (t_i) = \vq_i$, $\vq (t_f) =
\vq_f$ and $\ddot{\vq}(t) = \va(t)$.  Such a $\vq(t)$ is explicit%
ly given in particular by the object $\vq\lf(t; [\va(t')], \vq_f,
t_f, \vq_i, t_i\rt)$, a function-valued functional whose definition
is,
\re{
\vq\lf(t; [\va(t')], \vq_f, t_f, \vq_i, t_i\rt)\eqdf
\vq_i + (t - t_i)[(\vq_f - \vq_i)/(t_f - t_i) + {\ty\int_{t_f}^tdt'\,
(t' - t_i)^{-2}\int_{t_i}^{t'}}dt''\,(t'' - t_i)\va(t'')].
}{4a}
This particular $\vq(t)$ clearly satisfies $\vq (t_i) = \vq_i$ and
$\vq (t_f) = \vq_f$, and its first derivative with respect to $t$ is
given by the function-valued functional,
\re{\ba{c}
\dot{\vq}\lf(t; [\va(t')], \vq_f, t_f, \vq_i, t_i\rt) = \\
\m{} \\
(\vq_f - \vq_i)/(t_f - t_i) + {\ty\int_{t_f}^tdt'\,
(t' - t_i)^{-2}\int_{t_i}^{t'}dt''\,(t'' - t_i)\va(t'') +
(t - t_i)^{-1}\int_{t_i}^t}dt'\,(t' - t_i)\va(t'), \\
\m{} \\
\ea}{4b}
from which we readily calculate that its second derivative with res%
pect to $t$ comes out to equal $\va(t)$.  Conversely, \e{any} contin%
uously twice-differentiable path $\vq(t)$ that is defined for
$t\in [t_i, t_f]$ and which satisfies the two endpoint constraints
$\vq(t_i) = \vq_i$ and $\vq(t_f) = \vq_f$ is \e{equal} to $\vq\lf(t;
[\ddot{\vq}(t')], \vq_f,t_f, \vq_i, t_i\rt)$.  This follows from the
fact that for any continuously twice-differentiable path $\vq(t)$ that
is defined for $t\in[t_i, t_f]$ the relation,
\re{\vq(t) = \vq\lf(t; [\ddot{\vq}(t')], \vq(t_f), t_f, \vq(t_i), t_i\rt),
}{4c}
is an \e{identity}, as can be straightforwardly, albeit tediously,
verified by the use of the definition in Eq.~(4a) to expand out its
right-hand side, followed by repeated integrations by parts and
applications of the fundamental theorem of the calculus.

Since the \e{identity} given by Eq.~(4c) is not widely known, we
briefly digress to indicate how it can be derived.  For the contin%
uously twice-differentiable path $\vq(t)$ that is defined for
$t\in[t_i, t_f]$, we form the \e{error remainder}
$\vR\lf(t; [\vq(t')], t_i, t_f\rt)$ with respect to its \e{linear
interpolation} from $t_i$ to $t_f$,
\re{\vR\lf(t; [\vq(t')], t_i, t_f\rt)\eqdf\vq(t) - \vq(t_i) -
     (t - t_i)(\vq(t_f) - \vq(t_i))/(t_f - t_i).
}{5a}
Because this error remainder \e{vanishes} at both $t = t_i$ and
$t = t_f$, we make the \e{Ansatz} that,
\re{\vR\lf(t; [\vq(t')], t_i, t_f\rt) = (t - t_i){\ty\int_{t_f}^t}dt'\,
    \vUp\lf(t'; [\vq(t'')], t_i, t_f\rt).    
}{5b}
Now Eq.~(5a) implies that,
\re{\ddot{\vR}\lf(t; [\vq(t')], t_i, t_f\rt) = \ddot{\vq}(t),
}{5c}
which, given the \e{Ansatz} of Eq.~(5b), implies that,
\re{(t - t_i)\dot{\vUp}\lf(t; [\vq(t')], t_i, t_f\rt) +
2\vUp\lf(t; [\vq(t')], t_i, t_f\rt) = \ddot{\vq}(t).
}{5d}
This is a first-order inhomogeneous linear differential equation
for $\vUp\lf(t; [\vq(t')], t_i, t_f\rt)$ whose general solution is,
\re{\vUp\lf(t; [\vq(t')], t_i, t_f\rt) =
    (t - t_i)^{-2}{\ty\int_{t_0}^t}dt'\,(t' - t_i)\ddot{\vq}(t').
}{5e}
Since $\vR\lf(t; [\vq(t')], t_i, t_f\rt)$ vanishes at $t = t_i$, it is 
seen from the \e{Ansatz} of Eq.~(5b) that the only correct choice for
the unknown constant $t_0$ in Eq.~(5e) is $t_i$.  With that, Eq.~(4c)
is obtained from Eqs.~(5e), (5b) and (5a), together with the defini%
tion given in Eq.~(4a).  The relation of the Eq.~(4c) identity to the
error remainder of \e{linear interpolation} is quite analogous to the
relation of the identity,
\re{\vq(t) = \vq(t_0) + (t - t_0)\dot{\vq}(t_0) +
    {\ty\int_{t_0}^t dt'\int_{t_0}^{t'}}dt''\,\ddot{\vq}(t''),
}{6}
to the error remainder of \e{first-order Taylor expansion}.  (Note
that the Eq.~(6) identity follows from straightforward iteration of
the fundamental theorem of the calculus.)

With the above theorems regarding the function-valued functional
$\vq\lf(t; [\va(t')], \vq_f, t_f, \vq_i, t_i\rt)$ that is defined by
Eq.~(4a) in hand, we can now efficaciously integrate over all continu%
ously twice-differentiable configuration paths $\vq(t)$ that are de%
fined for $t\in[t_i, t_f]$ and which adhere to the two endpoint con%
straints $\vq(t_i) = \vq_i$ and $\vq(t_f) = \vq_f$.  It is clear that
this is done by replacing all occurrences of $\vq(t)$ in the path in%
tegrand of Eq.~(2) by $\vq\lf(t; [\va(t')], \vq_f, t_f, \vq_i,
t_i\rt)$, followed by functionally integrating \e{without restriction}
over \e{all} continuous accelerations $\va(t)$ that are defined for
$t\in[t_i, t_f]$.  Therefore the merely schematic phase-space path in%
tegral of Eq.~(2) is mathematically joined to the understanding given
below it concerning the $\vq(t)$ configuration-path endpoint con%
straints by rewriting it as the following \e{unconstrained functional
integral},
\re{K_H(\vq_f, t_f; \vq_i, t_i) =
{\ty\int}\mathcal{D}^{(t\in [t_i, t_f])}_{[\va(t), \vp(t)]}\,
\exp\lf(iS_H\lf(\lf[\vq\lf(t; [\va(t')], \vq_f, t_f, \vq_i, t_i\rt),
\vp(t)\rt]; t_f, t_i\rt)/\h\rt),
}{7}
where the functional integration embraces \e{all} continuous $\va(t)$
and \e{all} continuously twice-differentiable $\vp(t)$ that are de%
fined for $t\in[t_i, t_f]$.  To make futher progress with the path in%
tegral $K_H(\vq_f, t_f; \vq_i, t_i)$ described by  Eq.~(7), we now
need to address the question of how to \e{actually carry out} its in%
dicated unconstrained integration over all the above-described func%
tions $(\va(t), \vp(t))$ that are defined for $t\in[t_i, t_f]$.

\subsubsection*{Normalized multiple integration over the
orthogonal components of functions}

The set of functions $(\va(t), \vp(t))$ defined for $t\in[t_i, t_f]$
that are described in connection with Eq.~(7) above comprises an
\e{infinite}-dimensional vector space.  Integration over any \e{fin%
ite}-dimensional vector space is, of course, routinely carried out as
normalized \e{multiple ordinary integration} over all values of the
\e{components} of the vectors of that space which arise from \e{any of
its complete orthogonal decompositions}.  Thus to integrate over the
space of N-dimensional vectors $\vX$, we simply perform an appropri%
ately normalized multiple integration over any of its complete sets of
$N$ mutually orthogonal components,
\[{\ty\int d^N\vX = M_N\int dX_1\int dX_2\ldots\int dX_N}.\]
Here $\vX=\Sigma_{k=1}^NX_k\vb_k$, where the $\vb_k$ comprise \e{any
complete set of} $N$ \e{mutually orthogonal basis vectors},
i.e., that satisfy $\vb_k\dt\vb_{k'}=0$ if $k\neq k'$.  Thus the $N$
multiple \e{integration variables} $X_k$ are the $N$ \e{orthogonal
expansion coefficients}, i.e. $X_k = \vb_k\dt\vX/\vb_k\dt\vb_k$.

Now the time interval $[t_i, t_f]$ \e{also} has \e{complete sets} of
real-valued, discrete \e{mutually orthogonal basis functions}
$B_k(t)$, $k=0,1,\ldots,K,\ldots\:$, that satisfy,
\[{\ty\int_{t_i}^{t_f}B_k(t)B_{k'}(t)dt = 0}\m{\ if $k\neq k'$.}\]
We can expand any of our functions $(\va(t), \vp(t))$ in such a com%
plete, real-valued, discrete orthogonal basis set,
\[(\va(t), \vp(t))={\ty\Sigma_{k=0}^{\infty}(\va_k,\vp_k)B_k(t)},\]
where the $(\va_k, \vp_k)$ are that function's \e{orthogonal expansion
coefficients}, i.e.,
\[(\va_k, \vp_k) = {\ty\int_{t_i}^{t_f}B_k(t)(\va (t),\vp (t))dt/
\int_{t_i}^{t_f}\lf(B_k(t)\rt)^2 dt}.\]
The integration in Eq.~(7) over \e{all} the functions $(\va(t),
\vp(t))$ is therefore an appropriately normalized multiple integration
over all the orthogonal expansion coefficients $(\va_k, \vp_k)$,
\re{
{\ty\int}\mathcal{D}^{(t\in [t_i, t_f])}_{[\va(t), \vp(t)]} =
\lim_{K\rta\infty}\:M_K
{\ty\int d^n\va_0\,d^n\vp_0 \int d^n\va_1\,d^n\vp_1\ldots
\int d^n\va_K\,d^n\vp_K,}
}{8}
where in the particular case of the Eq.~(7) functional integration
the \e{measure normalization factor} $M_K$ is determined by the
requirement of Eq.~(3b).

A very commonly invoked \e{slight variation} of the above \e{complete}
discrete orthogonal basis set approach to functional integration in%
volves a \e{sequence} of \e{incomplete} discrete \e{approximation} or%
thogonal basis sets to the intuitively appealing complete \e{continu%
um} orthogonal basis set of delta functions in time, $B_{t_c}(t)\eqdf
\dl(t-t_c)$, where $t_c\in[t_i,t_f]$.  Given a partition of the time
interval $[t_i,t_f]$ into $K+1$ disjoint time subintervals, where
$K=0,1,2,\ldots\;$, we can \e{approximate} $B_{t_c}(t)$ by
$B^K_{t_c}(t)$, which, for $t$ in any of the $K+1$ disjoint
time subintervals of $[t_i,t_f]$ equals the \e{inverse} of the
\e{duration of that time subinterval} when $t_c$ is \e{also} in
that subinterval, but equals zero otherwise.  Obviously there
are \e{only} $K+1$ \e{distinct} such approximating functions
$B^K_{t_c}(t)$, so we may define $B^K_k(t)\eqdf B^K_{t_c}(t)$,
where $t_c$ is \e{any} time element of time subinterval number
$k$, $k=0,1,\ldots,K$.  It is clear that $B^K_k(t)$ is
\e{orthogonal} to $B^K_{k'}(t)$ for $k\neq k'$.  One develops in
this way a \e{sequence} in $K$ of \e{incomplete} orthogonal basis
sets that each have \e{only} $K+1$ members.  When $K\rta\infty$,
the intuitively appealing \e{continuum} orthogonal basis set of delta
functions $B_{t_c}(t)=\dl(t-t_c)$, which is, of course,
\e{complete}, will (very nonuniformly) be recovered \e{provided}
that care is taken to ensure that the \e{durations} of \e{all} of
the \e{individual time subintervals} of partition number $K$ tend
toward \e{zero} in that limit.  Notwithstanding that this is the
intuitively appealing ``standard'' method of functional integration,
its highly \e{nonuniform} approach to the continuum delta function
orthogonal basis set via \e{sequentially disjoint} incomplete discrete
orthogonal basis sets could conceivably spawn convergence issues in
cases of unfavorably disposed functionals.

\subsubsection*{The momentum path integral}

In addition to the configuration path integral
$K_H(\vq_f, t_f; \vq_i, t_i)$ of Eq.~(7), which is based on the
classical action $S_H([\vq (t), \vp (t)]; t_f, t_i)$ of Eq.~(1a)
that is classically appropriate to endpoint constraints on the
\e{configuration} paths, there \e{also} exists a \e{momentum} path
integral which is based on the classical action
$S_H'([\vq (t), \vp (t)]; t_f, t_i)$ of Eq.~(1b) that is classically
appropriate to endpoint constraints on the \e{momentum} paths.  In
close analogy with Eq.~(2), the \e{schematic} form of this \e{momentum}
path integral is given by,
\re{K_H'(\vp_f, t_f; \vp_i, t_i) =
{\ty\int}\mathcal{D}^{(t\in [t_i, t_f])}_{[\vq (t), \vp (t)]}\,
\exp(iS_H'([\vq (t), \vp (t)]; t_f, t_i)/\h ),
}{9}
where the $(\vq (t), \vp (t))$ \e{phase-space paths} are continuously
twice differentiable, but \e{here} it is supposed to be \e{understood}
that it is the \e{momentum paths} $\vp(t)$ which are all restricted by
two endpoint constraints, namely that $\vp(t_i) = \vp_i$ and
$\vp(t_f) = \vp_f$.  Just as the configuration path integral
$K_H(\vq_f, t_f; \vq_i, t_i)$ is the time-evolution operator for the
wave function in configuration representation, the momentum path
integral $K_H'(\vp_f, t_f; \vp_i, t_i)$ is the time-evolution operator
for the wave function in momentum representation, so that we have, in
analogy to Eq.~(3a),
\re{\phi(\vp_f, t_f) =
{\ty\int}K_H'(\vp_f, t_f; \vp_i, t_i)\phi(\vp_i, t_i)d^n\vp_i,
}{10a}
which requires that,
\re{\lim_{t_f\rta t_i}K_H'(\vp_f, t_f; \vp_i, t_i) =
\dl^{(n)}\!(\vp_f - \vp_i).
}{10b}
\indent
In line with the discussion which follows Eq.~(4a), any continuously
twice-differentiable $\vp(t)$ defined for $t\in[t_i, t_f]$ that
satisfies the two endpoint constraints $\vp (t_i) = \vp_i$ and
$\vp (t_f) = \vp_f$ has the explicit representation
$\vp\lf(t; [\vw(t')], \vp_f, t_f, \vp_i, t_i\rt)$ whose definition is,
\re{\vp\lf(t; [\vw(t')], \vp_f, t_f, \vp_i, t_i\rt)\eqdf
\vp_i + (t - t_i)[(\vp_f - \vp_i)/(t_f - t_i) + {\ty\int_{t_f}^tdt'\,
(t' - t_i)^{-2}\int_{t_i}^{t'}}dt''\,(t'' - t_i)\vw(t'')],
}{11}
where the ``force-rate'' $\vw(t)$ is an unconstrained continuous
function that satisfies,
\[\vw(t) = \ddot\vp\lf(t; [\vw(t')], \vp_f, t_f, \vp_i, t_i\rt).\]
Therefore we now
follow the example set by the replacement of the schematic Eq.~(2)
by the explicit Eq.~(7) by replacing the schematic presentation of
$K_H'(\vp_f, t_f; \vp_i, t_i)$ given in Eq.~(9) by the explicit,
\re{K_H'(\vp_f, t_f; \vp_i, t_i) =
{\ty\int}\mathcal{D}^{(t\in [t_i, t_f])}_{[\vq (t), \vw (t)]}\,
\exp\lf(iS_H'\lf(\lf[\vq(t),
\vp\lf(t; [\vw(t')], \vp_f, t_f, \vp_i, t_i\rt)\rt]; t_f, t_i\rt)/\h\rt),
}{12}
which is an unconstrained \e{functional integral} over $(\vq(t),
\vw(t))$ that uses the function-valued functional defined by Eq.~(11)
\e{to explicitly incorporate the two endpoint constraints} which
are required of all the \e{permitted} momentum paths $\vp(t)$ by the
\e{supplementary words} that are given \e{below} the schematic Eq.~%
(9).  The measure normalization factor for the functional integration
of Eq.~(12) will, of course, be determined by the requirement of
Eq.~(10b).

We note that the structure of the momentum path integral
$K_H'(\vp_f, t_f; \vp_i, t_i)$ which is given by Eq.~(12) is highly
analogous to that of the configuration path integral
$K_H(\vq_f, t_f; \vq_i, t_i)$ as given by Eq.~(7).  Therefore the steps
of any derivation concerning $K_H'(\vp_f, t_f; \vp_i, t_i)$ which flows
from Eq.~(12) will obviously proceed in close parallel with the steps of
a \e{corresponding} derivation concerning $K_H(\vq_f, t_f; \vq_i, t_i)$ 
which flows from Eq.~(7).  For that reason we shall in the remainder of
this paper be pointing out only steps for derivations concerning
$K_H(\vq_f, t_f; \vq_i, t_i)$ which flow from Eq.~(7), and simply leave
the analogous steps for the corresponding derivations concerning the
momentum path integral $K_H'(\vp_f, t_f; \vp_i, t_i)$ which flow from
Eq.~(12) as straightforward exercises for the interested reader.

\subsection*{Path integral evaluation through first order in the
elapsed time}

From Eq.~(3b) it is apparent that we already know the value of the con%
figuration path integral $K_H(\vq_f, t_f; \vq_i, t_i)$ through \e{zeroth
order} in the elapsed time $\dlt\eqdf (t_f - t_i)$.  If we can
\e{extend} the evaluation of $K_H(\vq_f, t_f; \vq_i, t_i)$ through
\e{first order} in $\dlt$, that result, together with Eq.~(3a), will
yield a \e{first-order differential equation in time} for the quantum-%
mechanical wave function $\psi(\vq, t)$ in configuration representation.
Obtaining the \e{solution} of that differential equation in time
when the wave function has the \e{initial} value $\psi(\vq, t_i)$ at
time $t_i$ is \e{equivalent} to Eq.~(3a) itself---i.e., that solution
of the differential equation in time \e{reproduces} the effect of
applying the \e{full} path integral $K_H(\vq_f, t_f; \vq_i, t_i)$ to
$\psi(\vq, t_i)$, the initial value of the wave function at time $t_i$.
In other words, the evaluation of the path integral
$K_H(\vq_f, t_f; \vq_i, t_i)$ through \e{just first order} in the
elapsed time $\dlt$ yields a first-order in time differential equation
for the wave function whose solution \e{duplicates} evaluation of the
effect of the \e{full} path integral on an initial wave-function value.
This permits  evaluation of the \e{full} path integral
$K_H(\vq_f, t_f; \vq_i, t_i)$ to, in principle, be sidestepped, which
is a considerable incentive to work it out through \e{just first order}
in the elapsed time $\dlt$.

To carry out the evaluation of $K_H(\vq_f, t_f; \vq_i, t_i)$ through
just first order in $\dlt$, we will obviously first need to expand out
its \e{integrand} functional, namely the integrand of the functional
integral on the right-hand side of Eq.~(7), through first order in
$\dlt$.  That Eq.~(7) integrand functional is, of course, given by,
\re{
I_H([\va(t), \vp(t)]; \vq_f, t_f, \vq_i, t_i)\eqdf 
\exp\lf(iS_H\lf(\lf[\vq\lf(t; [\va(t')], \vq_f, t_f, \vq_i, t_i\rt),
\vp(t)\rt]; t_f, t_i\rt)/\h\rt).
}{13a}
We can express the above Eq.~(7) integrand functional in greater
detail by applying Eq.~(1a) to it,
\re{
I_H([\va(t), \vp(t)]; \vq_f, t_f, \vq_i, t_i) =
e^{i\int_{t_i}^{t_f}
\lf(\dot\vq\lf(t; [\va(t')], \vq_f, t_f, \vq_i, t_i\rt)\dt\vp(t) -
H\lf(\vq\lf(t; [\va(t')], \vq_f, t_f, \vq_i, t_i\rt),\,\vp(t),\,t\rt)
\rt)dt/\h}.
}{13b}
The expansion of $I_H([\va(t), \vp(t)]; \vq_f, t_f, \vq_i, t_i)$ through
first order in $\dlt$ would obviously be facilitated if we could expand
out $(\va(t), \vp(t))$ for $t\in [t_i, t_f]$ in the familiar but
\e{nonorthogonal} Taylor-expansion monomial basis,
\[T_k(t)=(t-\hf(t_f+t_i))^k/k!,\: k=0,1,2\ldots\;.\]
Use of this Taylor-expansion basis isn't very feasible here, however,
because the straightforward approach to functional integration requires
that an \e{orthogonal} basis be utilized.  But an \e{orthogonal-polyno%
mial} basis $\{B_k(t),\: k=0,1,2,\ldots\}$, whose members have the
\e{same leading behavior as those of the Taylor-expansion monomial
basis}, and which as well \e{share the crucial property of being of
order} $O\!\!\lf((\dlt)^k\rt)$ \e{for} $t\in [t_i,t_f]$, is readily
constructed by systematic successive orthogonalization of the monomial
$T_k(t)$ in the interval $[t_i, t_f]$.  One thus obtains $B_0(t) =
T_0(t) = 1$ and,
\[B_k(t)=(t-\hf(t_f+t_i))^k/k!+{\ty\Sigma_{j=1}^k}c^{(j)}_k
(t-\hf(t_f+t_i))^{k-j}(\hf(t_f-t_i))^j\m{\ for $k=1,2,\ldots\;$,}\]
where the $k$ dimensionless constants $c^{(1)}_k,\ldots,c^{(k)}_k$
are determined by the $k$ \e{orthogonality} requirements that,
\[{\ty\int_{t_i}^{t_f}}(t-\hf(t_f+t_i))^{k'}B_k(t)dt=0
\m{\ for $k'=0,1,\ldots,k-1$.}\]
With \e{dimensionless} $c^{(j)}_k$, $j=1,2,\ldots,k$, it is clear
that $B_k(t)$ is of order $O\!\!\lf((\dlt)^k\rt)$ for $t\in
[t_i,t_f]$, as is foreshadowed above, and the above scheme for
$B_k(t)$ \e{does indeed} produce dimensionless $c^{(j)}_k$ because,
\[{\ty\int_{t_i}^{t_f}}(t-\hf(t_f+t_i))^N dt =
(\hf(t_f-t_i))^{N+1}(1+(-1)^{N})/(N+1).\]
It is convenient to note here that the first three $B_k(t)$ are,
explicitly,
\[\m{$B_0(t)=1$,\ \ $B_1(t)=(t-\hf(t_f+t_i))$,\ \ $B_2(t)=
(t-\hf(t_f+t_i))^2/2-(\hf(t_f-t_i))^2/6$,}\]
which again illustrates the key fact that $B_k(t)$ is of order
$O\!\!\lf((\dlt)^k\rt)$ for $t\in [t_i,t_f]$.  In view of their pro%
perties, we can recognize these orthogonal basis polynomials $B_k(t)$
as \e{scaled, translated Legendre polynomials which have Taylor-like
normalizations}.

The expansion in this orthogonal-polynomial basis of a continuously
twice-differentiable momentum path $\vp(t)$ that is defined for
$t\in [t_i, t_f]$ is,
\[\vp (t)={\ty\Sigma_{k=0}^{\infty}}\vp_k B_k(t),\]
where the momentum path's orthogonal expansion coefficients $\vp_k$,
$k=0,1,2,\ldots\;$, with respect to this basis are given by,
\[\vp_k={\ty\int_{t_i}^{t_f}B_k(t)\vp(t)dt/
\int_{t_i}^{t_f}(B_k(t))^2 dt}.\]

If this expansion is replaced by just its leading $k=0$ term, the
\e{error} made is obviously $(\vp(t) - \vp_0)$, which, in detail, is
$\lf(\vp(t) - \int_{t_i}^{t_f}\vp(t')dt'/(t_f - t_i)\rt)$.  If we now
adopt the following $[t_i, t_f]$-interval \e{mean-value notation} for
functions $g(t')$ of $t'$ in $[t_i, t_f]$,
\[\la g(t')\ra_{t'\in [t_i, t_f]}\eqdf
{\ty\int_{t_i}^{t_f}}g(t')dt'/(t_f - t_i),\]
then this error can be rewritten as
$\la(\vp(t) - \vp(t'))\ra_{t'\in [t_i, t_f]}$.  Since $\vp(t)$ is
\e{continuously differentiable} for $t\in [t_i, t_f]$, the fundamen%
tal theorem of the calculus tells us that for $t, t'\in [t_i, t_f]$,
we may replace $(\vp(t) - \vp(t'))$ by
$\int_{t'}^t\dot\vp(t'')dt''$.  With that we are able to conclude
that the error made by substituting for a continuously differentiable
momentum path $\vp(t)$ just its leading $k=0$ orthogonal-polynomial
expansion term $\vp_0$ can, for $t\in [t_i, t_f]$, be written,
\[\lf(\vp(t) - \vp_0\rt) =
\lf\la{\ty\int_{t'}^t}\dot\vp(t'')dt''\rt\ra_{t'\in [t_i, t_f]},\]
whose right-hand side is clearly of order $O(\dlt)$ when
$t\in [t_i, t_f]$---note the \e{key role} which the fact that $\vp(t)$
is \e{continuously differentiable} plays in this conclusion.

Now such a momentum path $\vp(t)$ is, in fact, continuously \e{twice}
differentiable, which leads us to suspect that an \e{even stronger}
theorem holds for the error which is made by substituting for $\vp(t)$
the \e{sum} of its leading \e{two} orthogonal-polynomial expansion
terms, which is $[\vp_0 + \vp_1(t - \hf(t_f + t_i))]$.  Indeed, the
fact that $\vp(t)$ is continuously \e{twice} differentiable implies
that this considerably more complicated error can, via \e{two succes%
sive integrations by parts}, followed by repeated regroupings and ap%
plications of the fundamental theorem of the calculus, be written for
$t\in [t_i, t_f]$ in the form,
\[\ba{c}
\lf(\vp(t) - [\vp_0 + \vp_1(t - \hf(t_f + t_i))]\rt) =
(3/2)\lf\la\int_{t'}^t dt''\lf\la\int_{t^{(3)}}^{t''}dt^{(4)}\ddot\vp(t^{(4)})
\rt\ra_{t^{(3)}\in [t_i, t_f]}\rt\ra_{t'\in [t_i, t_f]} \\
\m{} \\
-(1/4)\lf\la\int_{t'}^t dt''\lf(\int_{t_f}^{t''}dt^{(3)}\ddot\vp(t^{(3)}) +
\int_{t_i}^{t''}dt^{(3)}\ddot\vp(t^{(3)})\rt)\rt\ra_{t'\in [t_i, t_f]} \\
\m{} \\
-2(t - \hf(t_i + t_f))(t_f - t_i)^{-2}\lf\la(t' - \hf(t_i + t_f))^3
\ddot\vp(t')\rt\ra_{t'\in [t_i, t_f]},
\ea\]
whose right-hand side consists of \e{three} terms, \e{each} of which
is clearly of order $O\!\!\lf((\dlt)^2\rt)$ when $t\in [t_i, t_f]$.

Another entity which enters into the integrand functional
$I_H([\va(t), \vp(t)]; \vq_f, t_f, \vq_i, t_i)$ on the right-hand side
of Eq.~(13b) is the \e{time derivative of the configuration path func%
tional,} $\dot\vq\lf(t; [\va(t')], \vq_f, t_f, \vq_i, t_i\rt)$, which
is explicitly given by Eq.~(4b).  Because the acceleration function
$\va(t)$ is \e{continuous} for $t\in [t_i, t_f]$, Eq.~(4b) tells us
that the error made by substituting for
$\dot\vq\lf(t; [\va(t')], \vq_f, t_f, \vq_i,t_i\rt)$ the \e{constant
velocity} $(\vq_f - \vq_i)/(t_f - t_i)$ is of order $O(\dlt)$.  It is
important, however, to be aware of the fact that this constant veloci%
ty \e{itself} is of order $O\!\!\lf((\dlt)^{-1}\rt)$!  Therefore the
error made by \e{substituting for the composite object},
\[{\ty\int_{t_i}^{t_f}\dot\vq\lf(t; [\va(t')], \vq_f, t_f, \vq_i, t_i\rt)
   \dt\vp(t)dt},\]
that appears on the right-hand side of Eq.~(13b) the \e{approxima%
tion},
\[\lf\la(\vq_f - \vq_i)\dt[\vp_0 + \vp_1(t - \hf(t_f + t_i))]
  \rt\ra_{t\in[t_i,t_f]},\]
which fortuitously evaluates to simply $(\vq_f - \vq_i)\dt\vp_0$,
is clearly of order $O\!\!\lf((\dlt)^2\rt)$.  With that, a signifi%
cant contribution to the integrand functional $I_H([\va(t), \vp(t)];
\vq_f, t_f, \vq_i, t_i)$, as it appears on the right-hand side of
Eq.~(13b), has been evaluated and found to have the simple form
$(\vq_f - \vq_i)\dt\vp_0$ through first order in $\dlt$.  This partic%
ular contribution \e{itself} is obviously of order
$O\!\!\lf((\dlt)^0\rt)$.

By way of making further progress toward completing the task of evalu%
ating through first order in $\dlt$ the integrand functional
$I_H([\va(t), \vp(t)]; \vq_f, t_f, \vq_i, t_i)$ as it appears on the
right-hand side of Eq.~(13b), we also note from the fact that the
the acceleration $\va(t)$ is continuous for $t\in [t_i, t_f]$, toge%
ther with the representation of {configuration path functional}
$\vq\lf(t; [\va(t')], \vq_f, t_f, \vq_i, t_i\rt)$ which is given by
Eq.~(4a), that the error made by substituting for this entity the
\e{straight-line path} $\vq_i + (t - t_i)(\vq_f - \vq_i)/(t_f - t_i)$
when $t\in [t_i, t_f]$, is of order $O\!\!\lf((\dlt)^2\rt)$.  Of
course this straight-line path \e{itself} is of order
$O\!\!\lf((\dlt)^0\rt)$.  Therefore, assuming that all the \e{gradi%
ents} of the \e{classical Hamiltonian} $H(\vq, \vp, t)$ are \e{contin%
uous}, the error made by \e{substituting for the composite object},
\[{\ty\int_{t_i}^{t_f}}
H\lf(\vq\lf(t; [\va(t')], \vq_f, t_f, \vq_i, t_i\rt),\,\vp(t),\,t\rt)dt,
\]
that occurs on the right-hand side of Eq.~(13b) the \e{approximation},
\[{\ty\int_{t_i}^{t_f}}H(\vq_i + (t - t_i)(\vq_f - \vq_i)/(t_f - t_i),\,
\vp_0,\,t_i)dt,\]
which, on changing the integration variable from $t$ to the dimension%
less $\lm\eqdf(t-t_i)/(t_f - t_i)$, assumes the form,
\[(\dlt){\ty\int_0^1}H(\vq_i + \lm(\vq_f - \vq_i),\,\vp_0,\,t_i)d\lm,\]
is clearly of order $O\!\!\lf((\dlt)^2\rt)$.  With that, the key
\e{remaining} contribution to $I_H([\va(t), \vp(t)]; \vq_f, t_f,
\vq_i, t_i)$, as it appears on the right-hand side of Eq.~(13b), has
been evaluated through first order in $\dlt$.  As we see from its dis%
played form given just above, this particular contribution \e{itself}
is of order $O(\dlt)$.

With \e{both key contributions} through first order in $\dlt$ to
$I_H([\va(t), \vp(t)]; \vq_f, t_f, \vq_i, t_i)$ as it appears on the
right-hand side of Eq.~(13b) \e{now in hand}, we are finally in a
position to write down the result for
$I_H([\va(t), \vp(t)]; \vq_f, t_f, \vq_i, t_i)$ \e{itself} through
first order in $\dlt$,
\re{\ba{c}
I_H([\va(t), \vp(t)]; \vq_f, t_i + \dlt, \vq_i, t_i) = \\
\m{} \\
e^{i(\vq_f - \vq_i)\dt\vp_0/\h - i(\dlt/\h)
\int_0^1H(\vq_i + \lm(\vq_f - \vq_i),\,\vp_0,\,t_i))d\lm +
O\!\lf((\dlt)^2\rt)} = \\
\m{} \\
e^{i(\vq_f - \vq_i)\dt\vp_0/\h}\lf(1 - i(\dlt/\h)
\int_0^1H(\vq_i + \lm(\vq_f - \vq_i),\,\vp_0,\,t_i)d\lm\rt) +
O\!\!\lf((\dlt)^2\rt). \\
\ea}{14}
\indent
We see from Eqs.~(14) and (13a) that the integrand functional of the
functional integral on the right-hand side of Eq.~(7) has, through
first order in $\dlt$, \e{no} dependence on the acceleration func%
tion $\va(t)$ and \e{only} depends on the momentum path $\vp(t)$
through its $k=0$ orthogonal-polynomial expansion coefficient $\vp_0$.
This \e{independence through first order in} $\dlt$ of the \e{inte%
grand} of the functional integral on the right-hand side of Eq.~(7) of
\e{all but one} of the orthogonal-polynomial expansion coefficients
$\va_0, \vp_0, \va_1, \vp_1, \ldots, \va_K, \vp_K, \ldots\ $ of the
function $(\va(t), \vp(t))$ implies, via the multiple-integration pre%
scription given by Eq.~(8) for that functional integration, that we
have a \e{formally divergent}, undefined result for $K_H(\vq_f, t_i +
\dlt; \vq_i, t_i)$ through first order in $\dlt$!  The \e{cause} of
this predicament is that the Eq.~(14) expansion through first order
in $\dlt$ of the integrand functional of the functional integral on
the right-hand side of Eq.~(7), while true for \e{any given} function
$(\va(t), \vp(t))$, \e{loses its validity} when considered over the
\e{entire set} of such functions.  For example, over the \e{entire
set} of continuous acceleration functions $\va(t)$, the \e{error} made
in substituting for the time derivative of the configuration path
functional $\dot\vq\lf(t; [\va(t')], \vq_f, t_f, \vq_i, t_i\rt)$ of
Eq.~(4b) the constant velocity $(\vq_f - \vq_i)/(t_f - t_i)$ can obvi%
ously be made \e{arbitrarily large}, notwithstanding that it is clear%
ly of order $O(\dlt)$ for \e{any particular} continuous acceleration
function $\va(t)$.

Therefore the \e{only way} to deal with this quandary concerning the
functional integration of integrand functionals that are given
\e{through just a certain order of} $\dlt$ is to \e{cut off} the inte%
gration which \e{normally} runs over the \e{entire set} of applicable
functions.  From a physics point of view the imposition of such a
function integration cutoff does not seem very concerning because path
integrals are typically dominated by a \e{quite narrow range of func%
tions} whose actions differ by no more than several times $\h$ from
the action of the \e{classical} solution.  Nevertheless, the \e{possi%
bility} that the value of the thus cut-off functional integral
\e{might} irrevocably depend on the \e{details} of the cutoff which is
imposed is a dismaying one.  Fortunately, such cutoffs seem to typi%
cally affect only the value of an \e{overall factor} which multiplies
the \e{rest} of the functional integration result, and the path-inte%
gral requirement of Eq.~(3b) ensures that the measure normalization
factor $M_K$ present in the Eq.~(8) multiple-integration prescription
for the Eq.~(7) functional integration \e{cancels out} such factors.

Returning now to the Eq.~(14) result for the \e{integrand functional}
through first order in $\dlt$ of the functional integral which appears
on the right-hand side of Eq.~(7), we note that this integrand's
\e{lack of dependence} on the orthogonal-polynomial expansion coeffi%
cients $\va_0, \va_1, \vp_1, \va_2, \vp_2, \ldots, \va_K, \vp_K,
\ldots\,$ \e{leaves us no choice but to cut off} the Eq.~(8) multiple
integrations \e{over these particular coefficients}.  From this \e{un%
avoidably cutoff-modified} Eq.~(8) prescription for the functional in%
tegral of Eq.~(7) that applies to its integrand through just first or%
der in $\dlt$ as given by Eq.~(14), we obtain for $K_H(\vq_f, t_i +
\dlt; \vq_i, t_i)$ through first order in $\dlt$,
\re{\ba{c}
K_H(\vq_f, t_i + \dlt; \vq_i, t_i) = \\
\m{} \\
{\dy\lim_{K\rta\infty}}\:M_K\,F_K(A_0, A_1, P_1, A_2, P_2, \ldots,
A_K, P_K)\x \\
\m{} \\
\lf(\int d^n\vp_0\,e^{i(\vq_f - \vq_i)\dt\vp_0/\h}\lf(1 - i(\dlt/\h)
\int_0^1H(\vq_i + \lm(\vq_f - \vq_i),\,\vp_0,\,t_i)d\lm\rt)\rt) +
O\!\!\lf((\dlt)^2\rt),  \\
\ea}{15}
where the overall multiplicative factor
$F_K(A_0, A_1, P_1, A_2, P_2, \ldots, A_K, P_K)$ is the product of all
the \e{unavoidably cut-off integrals} over Eq.~(8) orthogonal-poly%
nomial expansion coefficients, namely,
\[\ba{c}
F_K(A_0, A_1, P_1, A_2, P_2, \ldots, A_K, P_K)\eqdf \\
\m{} \\
\int_{\{|\va_0|\le A_0\}} d^n\va_0
\int_{\{|\va_1|\le A_1, |\vp_1|\le P_1\}} d^n\va_1 d^n\vp_1
\int_{\{|\va_2|\le A_2, |\vp_2|\le P_2\}} d^n\va_2 d^n\vp_2\ldots
\int_{\{|\va_K|\le A_K, |\vp_K|\le P_K\}} d^n\va_K d^n\vp_K\,. \\
\ea\]
Now since,
\[{\ty\int}d^n\vp_0\,e^{i(\vq_f - \vq_i)\dt\vp_0/\h} =
(2\pi\h)^n\dl^{(n)}\!(\vq_f - \vq_i),\]
we must choose the Eq.~(8) measure normalization factor $M_K$ on the
right-hand side of Eq.~(15) to be equal to
$((2\pi\h)^nF_K(A_0, A_1, P_1, A_2, P_2, \ldots, A_K, P_K))^{-1}$ in
order to satisfy the $\dlt\rta 0$ path-integral limit requirement of
Eq.~(3b).  With that, Eq.~(15) becomes,
\re{
K_H(\vq_f, t_i + \dlt; \vq_i, t_i) =
\dl^{(n)}\!(\vq_f - \vq_i) - i(\dlt/\h)Q_H(t_i; \vq_f; \vq_i) +
O\!\!\lf((\dlt)^2\rt),
}{16a}
where,
\re{
Q_H(t_i; \vq_f; \vq_i)\eqdf 
{\ty\int_0^1d\lm\,(2\pi\h)^{-n}\int d^n\vp}\:
H(\vq_i + \lm (\vq_f - \vq_i), \vp , t_i)e^{i(\vq_f - \vq_i)\dt\vp /\h}.
}{16b}
It is easily demonstrated that $Q_H(t_i; \vq_f; \vq_i)$ is
\e{Hermitian}, i.e. that,
\re{
Q_H(t_i; \vq_f; \vq_i) = (Q_H(t_i; \vq_i; \vq_f))^{\ast}.
}{16c}
When Eq.~(16a) is combined with Eq.~(3a), we at long last obtain the
first-order differential equation in time that we have been seeking
for the wave function in configuration representation,
\re{
i\h\pa\psi(\vq_f, t_i)/\pa t_i =
{\ty\int}Q_H(t_i; \vq_f; \vq_i)\psi(\vq_i, t_i)d^n\vq_i,
}{17}
\m{}
\vspace{-2pc}

\subsection*{The quantized Hamiltonian operator and the Schr\"{o}dinger
             equation}

At this point we wish to mention the results for the momentum path
integral $K_H'(\vp_f, t_f; \vp_i, t_i)$ that parallel those which
we have just demonstrated for the configuration path integral
$K_H(\vq_f, t_f; \vq_i, t_i)$. Through first order in
$\dlt = (t_f -t_i)$, $K_H'(\vp_f, t_f; \vp_i, t_i)$ satisfies
relations which are highly analogous to those given in Eqs.~(16)
for $K_H(\vq_f, t_f; \vq_i, t_i)$, namely,
\re{
K_H'(\vp_f, t_i + \dlt; \vp_i, t_i) =
\dl^{(n)}\! (\vp_f - \vp_i) - i(\dlt/\h)Q_H'(t_i; \vp_f; \vp_i) +
O\!\!\lf((\dlt)^2\rt),
}{18a}
where,
\re{
Q_H'(t_i; \vp_f; \vp_i)\eqdf 
{\ty\int_0^1d\lm\,(2\pi\h)^{-n}\int d^n\vq}\:
H(\vq , \vp_i + \lm (\vp_f - \vp_i), t_i)e^{-i(\vp_f - \vp_i)\dt\vq /\h}.
}{18b}
It is easily demonstrated that $Q_H'(t_i; \vp_f; \vp_i)$ is
\e{Hermitian}, i.e. that,
\re{
Q_H'(t_i; \vp_f; \vp_i) = (Q_H'(t_i; \vp_i; \vp_f))^{\ast}.
}{18c}
When Eq.~(18a) is combined with Eq.~(10a), we obtain in analogy with
Eq.~(17) a first-order differential equation in time for the wave
function in momentum representation,
\re{
i\h\pa\phi(\vp_f, t_i)/\pa t_i =
{\ty\int}Q_H'(t_i; \vp_f; \vp_i)\phi(\vp_i, t_i)d^n\vp_i,
}{19}
A \e{crucial} relationship which holds between
$Q_H'(t_i; \vp_f; \vp_i)$ and $Q_H(t_i; \vq_f; \vq_i)$ is that,
\re{
{\ty\int}d^n\vp_fd^n\vp_i\,\langle\vq_f|\vp_f\rangle Q_H'(t_i;\vp_f;\vp_i)
\langle\vp_i|\vq_i\rangle = Q_H(t_i; \vq_f; \vq_i),
}{20}
where we have used the standard Dirac quantum mechanics notation
for the overlap amplitude between a configuration state and a mo%
mentum state, i.e., $\langle\vq |\vp\rangle = e^{i\vp\dt\vq /\h}/
(2\pi\h )^{n/2}$ and $\langle\vp |\vq\rangle =
(\langle\vq |\vp\rangle)^{\ast}$.  To carry out the verification
of Eq.~(20), it is useful to make the $d\lm$-integration that
arises from $Q_H'(t_i; \vp_f; \vp_i)$ via Eq.~(18b) the \e{outer%
most} integration, and then change integration variables from the
$(\vp_f, \vp_i)$ pair to the $\vp = \vp_i + \lm (\vp_f - \vp_i)$
and $\vp_- = (\vp_f - \vp_i)$ pair.  This variable transformation
has unit Jacobian, and the $d^n\vp_-$-integration will give rise
to a delta function which, in turn, permits the $d^n\vq $-integ%
ration that arises from $Q_H'(t_i; \vp_f; \vp_i)$ via Eq.~(18b)
to be carried out.  The upshot is to leave only the
$d^n\vp $-integration and the $d\lm$-integration, both of which
indeed occur in $Q_H(t_i; \vq_f; \vq_i)$, which is itself, of
course, the result being sought.  With this outline of the pro%
cedure, we leave the remaining straightforward details of veri%
fying Eq.~(20) to the reader.

Eq.~(20) shows that $Q_H(t_i; \vq_f; \vq_i)$ and
$Q_H'(t_i; \vp_f; \vp_i)$ are, respectively, the configuration
and momentum representations of the \e{very same quantum me%
chanical operator}, which we shall now denote as $\q H(t_i)$.
Thus, in the standard Dirac quantum mechanics notation,
\re{
Q_H(t_i; \vq_f; \vq_i) =  \langle\vq_f|\q H(t_i)|\vq_i\rangle ,
}{21a}
and,
\re{
Q_H'(t_i; \vp_f; \vp_i) =  \langle\vp_f|\q H(t_i)|\vp_i\rangle .
}{21b}
Eqs.~(16c) and (18c) of course show that this quantum mechanical
operator $\q H(t_i)$ is a \e{Hermitian} one.  Furthermore, if we
transcribe the first-order in time differential equation that ap%
pears in Eq.~(17) into standard Dirac quantum mechanics notation
by rewriting $\psi(\vq, t)$ as $\la\vq|\psi(t)\ra$ and switching
to the notation on the \e{right-hand side} of Eq.~(21a), we can
readily demonstrate that Eq.~(17) is equivalent to,
\re{
i\h\pa|\psi(t)\ra/\pa t = \q H(t)|\psi(t)\ra,
}{22}
which is the familiar Schr\"{o}dinger equation for the time 
evolution of the quantum state vector $|\psi(t)\ra$ under the
influence of the Hermitian Hamiltonian operator $\q H(t)$.
Here, however, this Hamiltonian operator $\q H(t)$ is \e{uniquely
determined}, via Eqs.~(21a) and (16b) (or alternatively via
Eqs.~(21b) and (18b)) by the \e{classical Hamiltonian function}
$H(\vq, \vp, t)$ for the physical system.  Inter alia, this
\e{crystallizes} the correspondence principle in a \e{very strong
form} indeed.

Of course one can tread a highly analogous route with the first-%
order in time differential equation that appears in Eq.~(19) by
rewriting $\phi(\vp, t)$ as $\la\vp|\phi(t)\ra$ and switching to
the notation on the right-hand side of Eq.~(21b), following which
it is readily demonstrated that Eq.~(19) is equivalent to,
\re{
i\h\pa|\phi(t)\ra/\pa t = \q H(t)|\phi(t)\ra,
}{23}
which is exactly the same quantum state vector Schr\"{o}dinger
equation as that of Eq.~(22).  Again, its controlling Hamiltonian
operator $\q H(t)$ is \e{uniquely determined}, via Eqs.~(21b) and
(18b) (or alternatively via Eqs.~(21a) and (16b)) by the \e{clas%
sical Hamiltonian function} $H(\vq, \vp, t)$ for the physical
system.

The \e{unique} classically underpinned Hamiltonian operator
$\q H(t_i)$ of Eqs.~(21), (16b) and (18b) was first obtained
from the Hamiltonian phase-space path integral by Kerner and Sut%
cliffe~\ct{K-S}, but it had been mooted by Born and Jordan~\ct{B-J}
in their pre-Dirac version of quantum mechanics.  Born and Jordan's
theory featured commutation rules which were more elaborate than
those of Dirac, but those rules were nevertheless still \e{not} suf%
ficiently strong to \e{uniquely} pin down the particular $\q H(t_i)$
of Eqs.~(21). Therefore Born and Jordan's discovery of $\q H(t_i)$
must be regarded as fascinatingly fortuitous rather than wholly sys%
tematic.  Dirac, with his Poisson bracket insight into quantum commu%
tators, had an excellent chance to \e{uniquely} pin down \e{exactly}
this $\q H(t_i)$, but ironically he ended up choosing commutation
rules that were even much \e{weaker}~\ct{Di2} than those of his pred%
ecessors Born and Jordan!  Kerner~\ct{Kr} was apparently the first to
work out the slightly strengthened self-consistent canonical commuta%
tion rule that Dirac \e{ought}, by rights, to have lit upon, but very
unfortunately Kerner failed to publish that work.  We shall briefly
develop the highly satisfactory canonical commutation rule that Dirac
\e{missed} in the last section of this paper.

\subsection*{Quantum amplitudes for individual configuration or
momentum paths}

As an extension of the interpretation that we have given to the con%
figuration path integrals of Eq.~(2) and Eq.~(7), it seems reasonable
to interpret the corresponding unconstrained functional integral over
\e{only momentum paths} $\vp(t)$, namely,
\re{
A_H([\vq (t)]; t_f, t_i)\eqdf 
{\ty\int}\mathcal{D}^{(t\in [t_i, t_f])}_{[\vp (t)]}\,
\exp(iS_H([\vq (t), \vp (t)]; t_f, t_i)/\h ),
}{24a}
as the quantum amplitude that the dynamical system traverses a speci%
fied \e{configuration path} $\vq(t)$ for $t\in [t_i, t_f]$.  If we now
also ponder the interpretation that we have given to the momentum path
integrals of Eq.~(9) and Eq.~(12), it as well seems reasonable that
the quantum amplitude that the dynamical system traverses a specified
\e{momentum path} $\vp(t)$ for $t\in [t_i, t_f]$ ought to similarly be
given by the corresponding unconstrained functional integral over
\e{only configuration paths} $\vq(t)$,
\re{
A_H'([\vp (t)]; t_f, t_i)\eqdf 
{\ty\int}\mathcal{D}^{(t\in [t_i, t_f])}_{[\vq (t)]}\,
\exp(iS_H'([\vq (t), \vp (t)]; t_f, t_i)/\h ).
}{24b}
\indent
Now we note from Eq.~(1a) that the \e{unconstrained} variation of the
classical action $S_H([\vq (t), \vp (t)]; t_f, t_i)$ with respect to the
\e{momentum path} $\vp (t)$ yields the \e{first} classical Hamiltonian
equation, and from Eq.~(1b) that the \e{unconstrained} variation of the
classical action $S_H'([\vq (t), \vp (t)]; t_f, t_i)$ with respect to
the \e{configuration path} $\vq (t)$ yields the \e{second} classical
Hamiltonian equation.  We therefore see that our above unconstrained
functional integrals in Eq.~(24a) for $A_H([\vq (t)]; t_f, t_i)$ and in
Eq.~(24b) for $A_H'([\vp (t)]; t_f, t_i)$ are the \e{precise embodiments}
of the principle that the \e{quantization of classical dynamics} is
achieved by substituting \e{superposition} of the exponential of $(i/\h)$
times the classical action for \e{variation} of that action.  (Additional%
ly, of course, that classical action must \e{not} be one that \e{implicit%
ly contravenes} the uncertainty principle!)  This \e{validates} the inter%
pretation of $A_H([\vq (t)]; t_f, t_i)$ as the \e{quantum amplitude}
that the dynamical system traverses the specified \e{configuration path}
$\vq (t)$ for $t \in [t_i, t_f]$ and of $A_H'([\vp (t)]; t_f, t_i)$ as the
\e{quantum amplitude} that the dynamical system traverses the
specified \e{momentum path} $\vp (t)$ for $t \in [t_i, t_f]$.  The \e{dom%
inant} stationary phase $\vp(t)$ \e{momentum} path that \e{contributes}
to $A_H([\vq (t)]; t_f, t_i)$ is readily seen to be the one that comes
from \e{algebraically} solving the \e{first} classical Hamiltonian equa%
tion, i.e.,
\re{
\dot\vq (t) = \del_{\!\vp (t)}H(\vq (t), \vp (t), t),
}{25a}
whereas the \e{dominant} stationary phase $\vq(t)$ \e{configuration} path
that \e{contributes} to $A_H'([\vp (t)]; t_f, t_i)$ is seen to be the one
that comes from \e{algebraically} solving the \e{second} classical Hamil%
tonian equation, i.e.,
\re{
\dot\vp (t) = -\del_{\!\vq (t)}H(\vq (t), \vp (t), t).
}{25b}
\indent
In order to obtain the configuration path integral
$K_H(\vq_f, t_f; \vq_i, t_i)$
described below Eq.~(2), we clearly must superpose the amplitudes
for all configuration paths $\vq(t)$ that satisfy the two \e{end%
point constraints} $\vq (t_i) = \vq_i$ and $\vq (t_f) = \vq_f$,
i.e., we must superpose $A_H([\vq(t)]; t_f, t_i)$ over all the
$\vq(t)$ which satisfy these two endpoint constraints.  A mathe%
matically efficacious method for superposing over only those
$\vq(t)$ which conform to these two endpoint constraints has al%
ready been developed with the aid of the configuration path func%
tional of Eq.~(4a), and application of this method to the Eq.~(24a)
representation of $A_H([\vq(t)]; t_f, t_i)$ will clearly produce
Eq.~(7).  Obtaining $K_H'(\vp_f, t_f; \vp_i, t_i)$ from the Eq.~(24b)
representation of $A_H'([\vp (t)]; t_f, t_i)$ with the aid of the
momentum path functional of Eq.~(11) proceeds along closely parallel
lines, and analogously produces Eq.~(12).

We note that in the mistaken Feynman-Dirac \e{Lagrangian}-action
version of the
configuration path integral $K_H(\vq_f, t_f; \vq_i, t_i)$, the ampli%
tude for the path $\vq(t)$, namely $A_H([\vq(t)]; t_f, t_i)$, is
\e{not} given by Eq.~(24a) but \e{instead} by the exponential of
$(i/\h)$ times the \e{Lagrangian} action for that configuration path
$\vq(t)$.  This \e{phase factor} is \e{only the integrand correspond%
ing to one particular momentum path} of the Eq.~(24a) \e{functional
integral} for $A_H([\vq (t)]; t_f, t_i)$ that runs over \e{all} momen%
tum paths.  The \e{particular} momentum path $\vp(t)$ which the Feyn%
man-Dirac Lagrangian-action hypothesis \e{inadvertently singles out}
is the one which the Lagrangian \e{implicitly determines} (in \e{con%
travention} of the uncertainty principle) from the configuration path
$\vq(t)$, namely,
\[\vp (t) = \del_{\!\dot\vq(t)} L(\dot\vq(t), \vq(t), t).\]
From classical mechanics one readily verifies that this \e{particular}
momentum path is in fact the \e{dominant} contributor to the \e{actu%
ally required sum} over momentum paths in Eq.~(24a) because it is
\e{precisely} the one which algebraically satisfies the Eq.~(25a)
first classical Hamiltonian equation.  That explains why the Lagran%
gian path integral ``works'' under certain favorable conditions, and
it \e{also} explains why, even under the \e{most favorable} of those
conditions, namely Hamiltonians which are quadratic forms in $\vp(t)$
whose Gaussian-phase functional integrals over the $\vp(t)$ \e{auto%
matically} produce the \e{dominant} phase factor, the Lagrangian path
integral \e{still} requires a mysterious \e{additional} factor---this
``mystery factor'' arises because \e{integration} over even \e{Gaus%
sian} phases yields \e{not only the dominant phase factor}, but a
\e{non-phase} factor as well. In the \e{subsequent} integration over
endpoint-constrained \e{configuration paths}, this factor is \e{not},
as Feynman's mistaken Lagrangian approach drove him to erroneously
conclude, a \e{totally ad hoc} measure ``normalizing factor'' which
needs to be puzzled out and \e{inserted by hand}, but rather \e{part
of the correct integrand}.  The Lagrangian path integral is thus seen
to be a not-quite-adequate \e{relative} of \e{systematic} semiclassi%
cal asymptotic approximations to the correct Hamiltonian phase-space
path integral.

\subsection*{The slightly stronger self-consistent canonical commutation
             rule Dirac missed}

The \e{unique} Hamiltonian quantization given by Eqs.~(21) in conjunc%
tion with Eq.~(16b) or Eq.~(18b) could \e{very well} have been discov%
ered by Dirac when he was formulating his canonical commutation rule
in 1925~\ct{Di2}, or at any time thereafter that he should have chosen
to \e{revisit} that work.  We now briefly explore just what it was
that Dirac \e{failed to light on} during an entire lifetime (see ref%
erence~\ct{Ka} for greater detail).  We note that the canonical commu%
tation rules which Dirac ended up postulating in 1925 (after some
struggling) can be gathered into the single formula,
\re{[c_1\id + \vk_1\dt\qvq + \vl_1\dt\qvp ,
     c_2\id + \vk_2\dt\qvq + \vl_2\dt\qvp] =
     i\h (\vk_1\dt\vl_2 - \vl_1\dt\vk_2)\id,
}{26a}
where $c_1$ and $c_2$ are constant scalars, and $\vk_1$, $\vl_1$,
$\vk_2$, $\vl_2$ are constant vectors.  The above equation can be
reexpressed in the much more suggestive form,
\re{[\Q{c_1 + \vk_1\dt\vq + \vl_1\dt\vp} ,
     \Q{c_2 + \vk_2\dt\vq + \vl_2\dt\vp}] =
     i\h\Q{ \{ c_1 + \vk_1\dt\vq + \vl_1\dt\vp ,
               c_2 + \vk_2\dt\vq + \vl_2\dt\vp \} }\, ,
}{26b}
where the overbrace denotes the \e{quantization} of the classical
dynamical variable beneath it, and the vertical curly brackets of
course denote the \e{classical} Poisson bracket.  (We use over%
braces to denote quantization \e{only} where the orthodox ``hat''
accent $\hat{}$, which is the standard way to denote quantization,
\e{fails} to be sufficiently \e{wide}.)  Eq.~(26b) is compellingly
elegant in light of Dirac's amazing groundbreaking demonstration
that the quantum mechanical analog of the classical Poisson brack%
et \e{must} be $(-i/\h )$ times the commutator bracket~\ct{Di2}.
Indeed, this form of Dirac's postulate rather strongly \e{suggests}
the possibility that it might perhaps be \e{strengthened} to simply
read,
\re{     [\Q{F_1(\vq , \vp )} , \Q{F_2(\vq , \vp )}] =
 i\h\Q{ \{   F_1(\vq , \vp )  ,    F_2(\vq , \vp )\} }.
}{27}
We note that the Eq.~(26b) form of Dirac's postulate is the
\e{restriction} of Eq.~(27) to $F_i(\vq , \vp )$, $i = 1, 2$,
that are both \e{inhomogeneous linear functions} of phase space.
Another, equivalent way to make this restriction is to require
that \e{all} the second-order partial derivatives of the
$F_i(\vq , \vp )$, $i = 1, 2$, \e{must vanish}.  Dirac was very
tempted by Eq.~(27), but upon playing with it he found to his
consternation that it \e{overdetermined} the quantization of
classical dynamical variables, and thus would be \e{self-incon%
sistent} as a postulate~\ct{Di2}.  Dismayed, he retreated to the
\e{restriction} on the $F_i(\vq , \vp )$ that results in Eq.~%
(26b), which, however, \e{cannot} determine the \e{order} of
noncommuting factors \e{at all}!  Far better that abject \e{un%
derdetermination} of the quantization of classical dynamical var%
iables than the outright \e{self-inconsistency} of their \e{over%
determination} was undoubtedly the thought that ran through
Dirac's mind.

But could there be a ``middle way'' that skirts overdetermination
\e{without} having to settle for \e{not} determining the order of
noncommuting factors \e{at all}?  Very unfortunately, Dirac ap%
parently \e{never} revisited this issue after 1925.  If one plays
with polynomial forms of the $F_i(\vq , \vp )$, one realizes that
the \e{overdetermination} does \e{not} occur if \e{no} monomials
that are dependent on \e{both} $\vq $ \e{and} $\vp $ are present.
This tells us that Dirac's restriction on the $F_i(\vq , \vp )$,
which requires that \e{all} their second-order partial deriva%
tives must vanish, is \e{excessive}: to prevent the self-incon%
sistent overdetermination of quantization it is \e{quite enough}
to require that \e{only} the \e{mixed} $\vq , \vp$ second-order
partial derivatives of the $F_i(\vq , \vp )$ must vanish, i.e.,
that,
\re{
     \del_{\vp}\del_{\vq}F_i(\vq , \vp ) = 0\m{, $i = 1, 2$},
}{28a}
which has the general solution,
$F_i(\vq , \vp ) = f_i(\vq ) + g_i(\vp )$, $i = 1 , 2$.  There%
fore, if we merely \e{replace} the Eq.~(26b) form of Dirac's
postulate by,
\re{
           [ \Q{f_1(\vq ) + g_1(\vp )} ,\Q{f_2(\vq ) + g_2(\vp )} ] =
   i\h\Q{ \{    f_1(\vq ) + g_1(\vp )  ,   f_2(\vq ) + g_2(\vp ) \} }\, ,
}{28b}
we will \e{still} have a canonical commutation rule that does
\e{not} provoke the self-inconsistent \e{overdetermination} of
classical dynamical variables.  But does it make any dent in
the gross \e{nondetermination} of the ordering of noncommuting
factors that characterizes Dirac's Eq.~(26b)?  The question of
whether a proposed approach fully determines the quantization
of \e{all} classical dynamical variables can be boiled down to
the issue of whether it fully determines the quantization of
the class of exponentials $\exp(i(\vk\dt\vq + \vl\dt\vp ))$,
because if it \e{does}, the \e{linearity} of quantization, com%
bined with Fourier expansion, then determines the quantization
of \e{all} dynamical variables.  It is apparent that
the only truly \e{new} consequence of Eq.~(28b) versus Dirac's
Eq.~(26b) is that,
\re{
    [ f(\qvq ), g(\qvp ) ] = i\h\Q{\del_{\vq}f(\vq )\dt\del_{\vp}g(\vp )}.
}{28c}
Putting now $f(\vq ) = e^{i\vk\dt\vq}$ and $g(\vp ) = e^{i\vl\dt\vp}$,
we see that Eq.~(28c) yields,
\re{
    \Q{e^{i(\vk\dt\vq + \vl\dt\vp )} } =
    (i/(\h\vk\dt\vl )) [ e^{i\vk\dt\qvq} , e^{i\vl\dt\qvp} ] ,
}{29}
which clearly answers the question concerning \e{full} determ%
ination of quantization in the affirmative!  It now remains to
be worked out how the unique, self-consistent quantization that
results from slightly \e{strengthening} Dirac's \e{excessively
restricted} canonical commutation rule of Eq.~(26b) to the
marginally \e{less} restricted canonical quantization rule of
Eq.~(28b) in fact \e{compares} with the unique quantization rule
of Eq.~(16b), which is a \e{key consequence} of the Hamiltonian
phase-space path integral.  To carry out the comparison, it is
very helpful to use the identity,
\re{
    [ e^{i\vk\dt\qvq} , e^{i\vl\dt\qvp} ] =
{\ty\int_0^1d\lm}\,\lf(d(e^{i\lm\vk\dt\qvq}e^{i\vl\dt\qvp}
 e^{i(1-\lm)\vk\dt\qvq})/d\lm\rt),
}{30a}
which is simply a consequence of the fundamental theorem of
the calculus.  Now if we carry out the differentiation under
the integral sign, there results,
\re{\ba{c}
    [ e^{i\vk\dt\qvq} , e^{i\vl\dt\qvp} ] =                            \\
\m{} \\
\int_0^1d\lm\, e^{i\lm\vk\dt\qvq}[i\vk\dt\qvq , e^{i\vl\dt\qvp}]
e^{i(1 - \lm )\vk\dt\qvq} = 
-i\h\vk\dt\vl\int_0^1d\lm\, e^{i\lm\vk\dt\qvq}e^{i\vl\dt\qvp}
e^{i(1 - \lm )\vk\dt\qvq} ,       \\
\m{}
\ea}{30b}
Combining this identity with the quantization result of Eq.~(29)
yields,
\re{
    \Q{e^{i(\vk\dt\vq + \vl\dt\vp )} } =
{\ty\int_0^1d\lm}\,e^{i\lm\vk\dt\qvq}e^{i\vl\dt\qvp}e^{i(1 - \lm )\vk\dt\qvq}. 
}{31}
We note here that the \e{form} of Eq.~(31) is that of a rule
for the \e{ordering} of noncommuting factors---and that rule
has a characteristically Born-Jordan~\ct{B-J}
\e{appearance}, i.e., \e{all} of the orderings of the class
that it embraces appear with \e{equal weight}.  H. Weyl, a
mathematician who liked to dabble in the new quantum mechan%
ics, thought it highly plausible that Nature would \e{se%
lect} the \e{most symmetric} of that class of orderings~\ct{We}%
, i.e., the one for which $\lm = \hf $, but Eq.~(31) has it
that Nature does \e{not select} amongst orderings \e{at all},
that it \e{instead} achieves an \e{alternate} kind of symmetry
through \e{utter nondiscrimination} amongst orderings (an
echo, perhaps, of the need to sum over \e{all} paths).  Now in
order to compare the quantization given by Eq.~(31) to the re%
sult of the integration which is called for by Eq.~(16b), we
must first obtain the \e{configuration representation} of the
former, which is facilitated by the well-known result that,
\de{ \langle\vq_f|e^{i\vl\dt\qvp}|\vq_i\rangle =
     \dl^{(n)}\! (\vq_f + \h\vl - \vq_i) .}
Using this, we obtain from Eq. (31) that,
\re{
    \langle\vq_f|\Q{e^{i(\vk\dt\vq + \vl\dt\vp )} }|\vq_i\rangle =
{\ty\int_0^1d\lm}\, e^{i\vk\dt (\vq_i + \lm (\vq_f - \vq_i))}\;
     \dl^{(n)}\! (\vq_f + \h\vl - \vq_i) ,
}{32}
which result, it is readily verified, is \e{also} produced
by the path integral quantization formula of Eq.~(16b)
when $e^{i(\vk\dt\vq + \vl\dt\vp )}$ is substituted for
the classical Hamiltonian.

We do not really need to go further than this to have
demonstrated that the quantization produced by the path
integral is the \e{same} as that produced by the slightly
strengthened canonical commutation rule of Eq.~(28b).  The
reader may find it interesting, however, to follow out
the full consequences of combining the \e{linearity} of
quantization with the \e{Fourier expansion} of an
\e{arbitrary} classical dynamical variable
$F(\vq , \vp )$, which together formally imply that,
\re{\ba{c}
    \langle\vq_f|\Q{F(\vq , \vp )}|\vq_i\rangle =                   \\
\m{} \\
(2\pi )^{-2n}\int d^n\vq 'd^n\vp '\, F(\vq ', \vp ')
\int d^n\vk d^n\vl\, e^{-i(\vk\dt\vq ' + \vl\dt\vp ')}
    \langle\vq_f|\Q{e^{i(\vk\dt\vq + \vl\dt\vp )} }|\vq_i\rangle .    \\
\m{}
\ea}{33a}
The next step is, of course, to substitute the unam%
biguous result for the quantization of the exponential
$e^{i(\vk\dt\vq + \vl\dt\vp )}$, which was obtained in
Eq.~(32) from the slightly strengthened canonical com%
mutation rule of Eq.~(28b), for the last factor of the
integrand on the right hand side of Eq.~(33a). We
leave it to the reader to then plow through all
the integrations that can be carried out in closed
form to obtain,
\re{
    \langle\vq_f|\Q{F(\vq , \vp )}|\vq_i\rangle = 
{\ty\int_0^1 d\lm\,(2\pi\h)^{-n}\int d^n\vp}\:
F(\vq_i + \lm (\vq_f - \vq_i), \vp )e^{i(\vq_f  - \vq_i)\dt\vp /\h},
}{33b}
which is precisely the \e{same} quantization result as
is obtained from the Hamiltonian phase-space path
integral, namely that given by Eq.~(16b), when
$F(\vq , \vp )$ is subtituted for the classical
Hamiltonian.  Dirac's 1925 postulation of Eqs.~(26)
as \e{the} canonical commutation rule is thus seen
to be a \e{purely historical aberration}.  One can only
suppose that if Dirac had \e{kept working} over the years
on trying to obtain a more satisfactory canonical commu%
tation rule than the abjectly deficient Eqs.~(26), he
would \e{surely} have eventually lit upon their slight
strengthening to Eq.~(28b), which \e{removes} their
vexing ordering ambiguity \e{without} imperiling their
self-consistency.  The Hamiltonian phase-space path
integral's utterly straightforward unique quantiza%
tion \e{ought} to have been the needed wake-up call
to the physics community on this issue, but by then
the result of Dirac's inadequate work had become so
\e{ingrained} that it was mentioned by Cohen~\ct{Co}
in his last paragraph as another reason to call
into question the correct path integral
results of Kerner and Sutcliffe~\ct{K-S}.
Cohen's mention of the ``usual'' ambiguity of
quantization may have been one of Kerner's motivations
to revisit Dirac's canonical commutation rule.  He soon
came up with its slight strengthening to Eq.~(28b) and
showed this to produce the very same Born-Jordan~\ct{B-J}
quantization as does the Hamiltonian phase-space path
integral~\ct{Kr}.  Stunningly, however, Kerner
\e{never published} those results!  Neither did he
\e{ever reply} in print nor at any scholarly forum
to the meritless $\lim_{|t_f-t_i|\rta 0}|\vq_f-\vq_i|=0$
objection that Cohen raised regarding his groundbreaking
paper with Sutcliffe on the consequences of the Hamiltonian
phase-space path integral.  Pressed on why, he said
that he ``did not want to pick a fight with Leon
Cohen''~\ct{Kr}.  Kerner's apparently shy, retiring nature
came within a hair of \e{denying} physics the gifts that
his mind had produced.  To read page after page of
solemn classification by Tirapegui et al.~\ct{Ti}
of wrong ``discretization'' results that flow from
Cohen's lapse is to utterly despair of Kerner's
choice of silence.
\pagebreak

\end{document}